# Nucleotide Distribution Patterns in Insect Genomes

Manoj Pratim Samanta
Systemix Institute, Los Altos, CA 94024
manoj.samanta@systemix.org

**This work analyzed genome-wide nucleotide distribution patterns in ten insect genomes. Two internal measures were applied – (i) GC variation and (ii) third codon nucleotide preference. Although the genome size and overall GC level did not show any correlation with insect order, the internal measures usually displayed higher levels of consistency. GC variations in genomes of hymenopteran insects, honeybee and wasp, ranked highest among all eukaryotic genomes analyzed by us. Genomes of honeybee and beetle, insects of different orders with similar overall GC levels, showed significant internal differences. Honeybee genome stood out as unusual due to its high GC variation and 'left-handed' gene locations.**

Recent sequencing of several insect genomes provides us with an unique opportunity to understand the nature of molecular level evolutionary forces and their manifestation in the phenotypic variations between the organisms [Adams-00, Holt-02, THGSC-06, Xia04, Zdobnov-02]. Much of the earlier genetic knowledge on insects were derived through a narrow prism of analysis of the *Drosophila* genome, but recent studies on other insects displayed additional unique characteristics. For example, the genome of hymenopteran insect honeybee had much lower GC level than the fly, isochore-like nucleotide distributions and preferences of genes to locate in the AT-rich segments of the genome [Elsik-07, Jorgensen-06, THGSC-06]. Among other insect genomes, GC level in coleopteran insect *Tribolium castaneum* was as low as honeybee, whereas another hymenopteran insect, *Nasonia* was GC-rich like *Drosophila*. These observations prompted us to perform a comparative study of nucleotide distributions in all insect genomes sequenced so far. Here we present the first report of this comparison. With availability of new data, these results will be continually updated at http://www.manojlabs.com/insects.

At present, sequencing of ten insect genomes are initiated or completed, and four others are also being considered (Table 1). Their phylogenetic relationship is shown in Fig. 1. All but two of them are from the endpoteran division. Among the endpoteran insects, the order Diptera is most well represented with eight members being considered for sequencing. From the other orders such as Hymenoptera, Coleoptera and Lepidoptera, two, one and one representatives are selected respectively. Among ten genomes, where sequencing has been initiated, seven are fully assembled and three are unassembled.

Following inferences can be drawn from a comparison of genome sizes and GC levels as shown in Table 1. The genome sizes vary widely from 130Mb in *Drosophila* to 1384 Mb in *Aedes*. Incidentally, both limits are represented by insects from the same order (Diptera). Large divergence is seen even within a suborder of dipterans, such as among two mosquitoes, *Anopheles* and *Aedes*. Overall GC levels for the assembled genome also vary widely from 32.7% in honeybee to 44.3% in mosquito *Anopheles*. In a separate study, we compared ten fungal genomes and observed a similar range, although, on average, the GC levels were ~10% higher in the fungal genomes (41-57%) ([http://www.manojlabs.com/genomes](http://www.manojlabs.com/genomes)). This work also made preliminary GC estimates for the unassembled insect genomes based on subsets of sequencing reads (see Methods). If those estimates truly reflect the GC levels of the overall genomes, the range of mean GC for insect and even Dipteran genomes would increase substantially, with *Culex* mosquito (63.2%) being twice as GC-rich as aphid (31.5%). We also note that both hymenopteran insects in Table 1 have similar genome sizes, but significantly different GC levels.

Above comparison demonstrates that the measures such as genome size and GC level do not maintain any identifiable pattern across the insect phylogenetic tree. This is not surprising, because random substitutions, insertions and deletions continually take place within the genomes, resulting in changes of genome lengths and GC compositions. Therefore, to derive additional insights into the insect genomes, this work applied two other measures – GC variation and third codon nucleotide composition - referred to as 'internal measures' from here on.  Similar measures were applied by Bernardi *et al.* to compare nucleotide distributions in vertebrates [Bernardi-05]. Following thoughts were applied in selecting those measures. Firstly, they are likely to be substantially unaffected by random substitutions, insertions and deletions, unless those random processes preferentially affect certain regions of the genomes than others. Secondly, these measures can be applied to early sequencing versions of the genomes, even prior to initial draft assemblies. In contrast, measures such as domain lengths [THGSC-06] cannot be correctly computed, unless a good quality assembly of the genome is available. Thirdly, these measures are easy to compute and interpret. This is an important factor, because similar calculations will be repeated for hundreds of genomes that are being sequenced or revised (http://www.manojlabs.com/genomes).

The first measure, genome-wide GC variation, was computed for each genome in the following manner. From an entire genome, 5,000 segments of length 1Kbases were randomly selected and their GC-levels were calculated. GC variation was defined as the standard deviation of those 5,000 samples (see Methods). GC variations for all assembled and unassembled insect genomes are shown in Table 1 and the distributions of 5,000 samples are plotted as histograms in Fig. 2. It can be shown that random mutation, even though it is biased to modify the overall GC level of a genome, should not significantly affect the GC variation or the shape of the distribution from one insect to another.

Following observations are made from comparison of GC variations among ten insect genomes. In all except the hymenopteran insects, the variations range between 5-7%, and are generally higher than the fungal genomes (3-4.5%). However, the variations in two hymenopteran insects (honeybee – 9.9%, *Nasonia* – 8.14%) are larger than all other eukaryotic genomes analyzed by us (e.g. human – 7.8%, sea urchin – 4.3%). Large variation in honeybee is unrelated to its general AT-richness, because *Tribolium* and aphid genomes with similar AT-levels do not show similar high variations. The manifestation of high variation in the sequence data is shown in Fig. 3, where GC levels in 6Mb regions of the bee and fly genomes are compared. Higher fluctuation in the local GC levels of bee genome is apparent from visual inspection of the images.

Further understanding of GC variations can be made from the distribution plots shown in Fig. 2. The same figure also includes the distributions of human and sea urchin genomes for comparison. Most genomes, including human, show unimodal Gaussian-like distributions with a long tail on the right. However, the distributions of the hymenopteran genomes show distinct broad patterns suggesting larger standard deviation or GC variation. Interestingly, distribution of the *Nasonia* genome is bimodal unlike other genomes. In an independent investigation, Jorgensen *et al.* reported observing bimodality in the distribution of third codons in honeybee genes, but such bimodality do not show in our dataset possibly due to sampling of the entire genome. Apart from Hymenoptera, unique pattern is seen in the distribution of the mosquito *Culex pipiens*. It is right-shifted (GC-rich) and inverted with large tail on the left. As a cautionary note, we mention that the data for this genome was derived from a subset of unassembled reads. Clearer picture will emerge, once the genome is fully assembled.

As a second internal measure, GC levels of third codons (GC3Cs) of protein-coding genes are compared with the rest of the genome. Median of GC3Cs of all coding genes is computed for every genome and drawn as red vertical lines in Fig. 2 along with the previously discussed distributions. Nearly 75% of third codons are neutral bases that could be either A/T or G/C. In unicellular organisms, such as yeast *S. cerevisiae* and diatom *T. pseudonana,* the red line of Fig. 2 nearly coincides with the mode of the distribution for the entire genome. Secondly, any evolutionary process of random mutation that raises or lowers the overall GC level of a genome should not affect the third codons of protein coding genes differently. Another significance of third codon nucleotide distribution is that in honeybee, it generally reflects the GC level of the genomic region where the gene is located [Jorgensen-07]. However, we do not know whether the same is true for other insects. Overall, any difference between the mode of the distribution for the entire genome and the third codons provides useful information regarding codon preferences of the protein-coding genes, and maybe, their locational preferences.

We computed the median GC3C measure for six assembled insect genomes, as well as for sea urchin and human genomes. For all genomes in Fig. 2 showing regular Gaussian-like distributions with long tails on the right, third codon distributions of coding genes are on the long tails. We name this pattern 'right-handed'. *Nasonia* genome also shows right-handed pattern, but the median GC3C is closer to the mode than others. A truly unique 'left-handed' pattern is seen in the bee genome. We note that among over 20 eukaryotic genomes analyzed by us, bee genome is the only one showing such clear 'left-handed' pattern. Yeast *S. cerevisiae* shows slightly 'left-handed' pattern, but the difference between the mode of the distribution and median GC3C is statistically insignificant (.02%). If the median GC3C in *Culex* mosquito is on the long tail like other genomes, it may also show similar 'left-handed' gene locations due to its inverted overall distribution. We note that extreme GC-richness of *Culex* genome does not require it to be 'left-handed'. The genome of *Chlamydomonas reinhardtii* is equally GC-rich (63%), but still remains right-handed with extremely high median GC3C (82%).

Fig. 4 compares the GC levels of entire coding regions and third codons of all genes separately for each genome. It is interesting that despite all differences between the genomes discussed earlier, broad features of this measure appear nearly identical (linear correlation with same slope) in all insects. Fig. 5 presents enlarged versions of the same plots for fly and bee. They show information similar to what has been discussed previously [THGSC-06] and in this report. In bee, the dots are clustered near the bottom, because the protein-coding genes prefer to locate in the AT-rich regions. The pattern is opposite in fly with the dots clustered near the top of the plot (GC-rich). Additional analysis will be presented in a future report.

What are the evolutionary implications of the observations discussed in this report? Or framing the question differently – what was the genome of the common ancestor of all insects like, and how did it evolve to create the diversity of distributions and gene locations as presented above? Nearly all insect genomes as well as human and sea urchin genomes show Gaussian-like distributions with longer tails on the right, and therefore it is likely to be the internal pattern for their common ancestor. This suggests that the honeybee genome evolved differently to emerge into a different internal configuration. If that is true, the followup question would be whether this process started in the common ancestor of all hymenopteran organisms, or whether it happened at a later time point. Similarity between the distributions of bee and *Nasonia* genomes, and their higher GC variations suggest that both were affected by the same processes. However, this does not explain the left- and right-handedness of median GC3Cs in bee and *Nasonia* respectively.

One hypothesis that may explain the above is as follows. Overall the data suggests the presence of multiple types of evolutionary processes of different timescale. A process of shorter time-scale that changes the GC levels of the genomes without affecting the internal measures is present in all genomes. This process likely constitutes of random substitutions, insertions and deletions. It is present in all insect genomes and tends to increase the GC levels of the protein-coding genes compared to rest of the genome. However, an additional longer timescale process must have been active in honeybee to cause its unusual distribution, including left-handedness in the bee genome. The common ancestor of the hymenopteran insects went through some unusual changes comprising of indiscriminate modification of all neutral bases to A/T. Further evidence of this presence of such forces will be presented in a different report [Samanta-07]. It lowered the GC level of the entire genome and also resulted in the left-handedness. In honeybee, this force is dominant over any other shorter time frame processes. A balance was achieved in *Nasonia* between this force and the shorter time frame process. Hence the genes were pushed to higher GC levels. Competition between two forces resulted in its bimodal distribution.

The honeybee genome paper mentioned: "Consistent with an (A+T)-rich genome, honeybee genes occur more frequently in (A+T)-rich domains compared with other species" [THGSC-06]. This analysis suggests that the honeybee genome is unique among the AT-rich genomes to have such preferences in gene locations. Also, the AT-rich isochores in honeybee are not merely mirror images of the GC-rich isochores in the mammalian genomes and may have originated from different processes. Apart from the use of the same word 'isochore', their evolutionary origin could be distinct. The pattern seen in honeybee is likely to be unique to the hymenopteran genomes.

In conclusion, this work compared all available insect genomes based on two easy-to-compute internal measures. Two hymenopteran genomes show most distinct patterns suggesting the presence of unexplained evolutionary forces in the formation of their common ancestor. In particular, the honeybee genome showed internal patterns different from all insect and non-insect eukaryotic genomes analyzed by us. Finally, an online resource is developed at http://www.manojlabs.com/genomes to present similar internal measures for all other sequenced eukaryotic genomes, such as fungi and chordates.

# Methods

**Availability of genomic data.**

Assembled genomes of *Bombyx mori* (version SW_scaffold_ge2k from silkDB)*, Tribolium castaneum* (version Tcas20051011 from Baylor HGSC)*, Apis mellifera* (V4 from Baylor HGSC)*, Nasonia vitripennis* (V0.5 from Baylor HGSC)*, Drosophila melanogaster* (V4.2.1 from flybase)*, Anopheles gambiae* (version agamP3.fa from vectorbase) and *Aedes aegypti* (version AEDES1 from vectorbase) were downloaded from their respective sequencing centers or insect-related databases. Web links for sequence sources are listed in http://www.manojlabs.com/genomes. For *Acyrthosiphon pisum, Culex pipiens* and *Ixodes scapularis,* unassembled reads were downloaded from either the NCBI trace archive (ftp://ftp.ncbi.nih.gov/pub/TraceDB/), or the Vectorbase website (http://www.vectorbase.org).

Gene predictions were available for *Tribolium castaneum* (GLEAN3 from Baylor), *Apis mellifera* (GLEAN3 from Baylor), *Drosophila melanogaster* (dmel-all-CDS-r4.2.1.fasta from flybase), *Anopheles gambiae* (agambiae.CDNA-KNOWN.AgamP3.3.fa from vectorbase) and *Aedes aegypti* (aaegypti.TRANSCRIPTS-AaegL1.1.fa from vectorbase). For *Nasonia vitripennis*, a preliminary set of predictions were derived by performing homology search of honeybee proteins on to V0.5 genome of *Nasonia*, and then considering only the longer exons.

**Definition and calculation of variation.**

For each inset, 5,000 segments of length 1 Kbases were selected randomly from the entire genome and their GC levels were calculated. Variation for a genome (Table 1) was defined as the standard deviation of those 5,000 GC-levels. The above calculations were improved with the following adjustments, where appropriate. If any chosen segment had over 50% of unsequenced bases, it was discarded and a new segment was chosen. For insects with unassembled genomes, 5,000 sequencing reads longer than 1 Kbases were randomly selected, and 1Kbase segment was randomly selected from each read.

# Acknowledgments

This report is dedicated to Prof. Gene Robinson for getting me curious about the nucleotide distribution in honeybee. The efforts by the respective sequencing centers in providing raw or assembled sequences for *Acyrthosiphon pisum, Aedes aegypti*, *Culex pipiens*, *Ixodes scapularis, Nasonia vitripennis* and *Tribolium castaneum* is duly acknowledged. We also thank Baylor Human Genome Sequencing Center for providing the predicted genes for *Tribolium castaneum*.

# References

[Adams-00] "The Genome Sequence of *Drosophila melanogaster*", M. D. Adams *et. al., Science* **287**, 2185-2195 (2000).

[Bernardi-05] Structural and Evolutionary Genomics, Volume 37: Natural Selection in Genome Evolution (New Comprehensive Biochemistry) Giorgio Bernardi, *Elsevier Science* (2005) .

[Elsik-07] C. Elsik *et al.* (unpublished).

[Holt-02] "The Genome Sequence of the Malaria Mosquito *Anopheles gambiae*", R. A. Holt *et al.*, *Science* **298**, 129 (2002).

[Jorgensen-06] "Heterogeneity in regional GC content and differential usage of codons and amino acids in GC-poor and GC-rich regions of the genome of *Apis mellifera*", F. G. Jorgensen, M. H. Schierup, A. G. Clark, *Molecular Biology and Evolution,* **Online-Dec (2006)**.

[Samanta-07] M. P. Samanta, *Systemix Reports* 3 (unpublished).

[THGSC-06] "Insights into social insects from the genome of the honeybee *Apis mellifera*", The Honeybee Genome Sequencing Consortium, *Nature* **443**, 932-949 (2006).

[Xia-04] "A draft sequence for the genome of the domesticated silkworm (*Bombyx mori*)", Q. Xia *et al., Science* **306**, 1937-40 (2004).

[Zdobnov-02] "Comparative genome and proteome analysis of *Anopheles gambiae* and

*Drosophila melanogaster*", E. M. Zdobnov *et al. Science* **298**, 149–159 (2002).

# Tables

| *Organism* | *Common name* | *Division (Order)* | *Size of assembled genome (Mb)* | *GC-level* | *GC Variation* |
|---|---|---|---|---|---|
| *Bombyx mori* | silkworm | Endoptera (Lepidoptera) | 174 (ge2k) | 37.40% | 5.50% |
| *Tribolium castaneum* | beetle | Endoptera (Coleoptera) | 199.6 (V0.5) | 33.90% | 6.60% |
| *Apis mellifera* | honeybee | Endoptera (Hymenoptera) | 235 (V4) | 32.70% | 9.90% |
| *Nasonia vitripennis* | wasp | Endoptera (Hymenoptera) | 234 (V0.5) | 41.80% | 8.14% |
| *Drosophila melanogaster* | fruitfly | Endoptera (Diptera) | 129.9 (V4.2.1) | 42.30% | 5.66% |
| *Anopheles gambiae* | mosquito | Endoptera (Diptera) | 278 (AgamP3) | 44.30% | 6.56% |
| *Aedes aegypti* | mosquito | Endoptera (Diptera) | 1384 (AEDES1) | 38.10% | 5.64% |
| *Culex pipiens* | mosquito | Endoptera (Diptera) | Unassembled | 63.20% | 6.94% |
| *Acyrthosiphon pisum* | aphid | Endoptera (Diptera) | Unassembled | 31.50% | 6.39% |
| *Ixodes scapularis* | deer tick | Hemiptera | Unassembled | 46.40% | 5.92% |
| *Rhodnius prolixus* | bug-Chagas' vector | Hemiptera | Initiated | X | X |

| *Organism* | *Common name* | *Division (Order)* | *Size of assembled genome (Mb)* | *GC-level* | *GC Variation* |
|---|---|---|---|---|---|
| *Phlebotomus papatasi* | sandfly | Endoptera (Diptera) | Initiated | X | X |
| *Glossina morsitans* | testesfly | Endoptera (Diptera) | Initiated | X | X |
| *Lutzomyia longipalpis* | sandfly | Endoptera (Diptera) | Initiated | X | X |

**Table 1. Insect genomes.** Sequencing of fourteen insect genomes have been considered so far. Among them, ten were initiated and seven already assembled. Their phylogenetic relationship is shown in Figure 1. For the completed genomes, overall GC levels and variations are shown. Preliminary estimates of GC levels for the unassembled genomes are made from subsets of sequencing reads (see Methods). Similar data for a large number of eukaryotic genomes are presented in **http://www.manojlabs.com/genomes** and will be continually updated with sequencing of new genomes.

# Figures

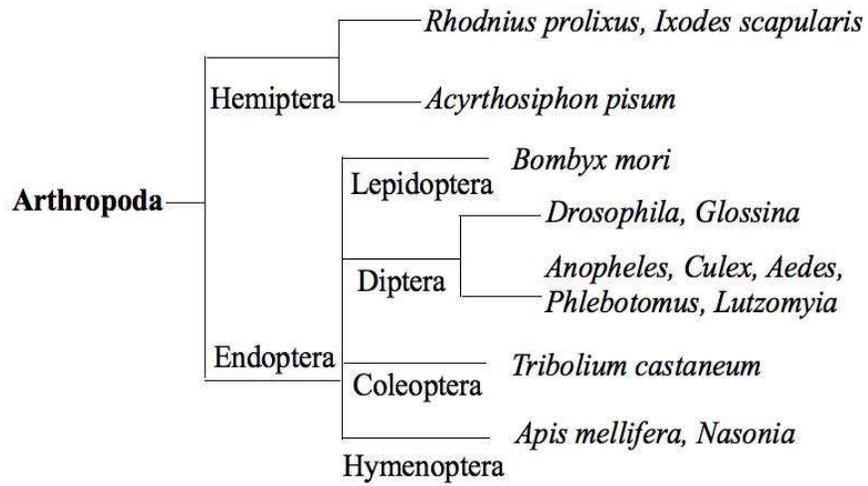

**Figure 1. Phylogenetic relationship between the sequenced insects.** Sequencing of fourteen insect genomes were initiated so far. Dipteran order is most well-represented, because those insects are typical disease vectors.

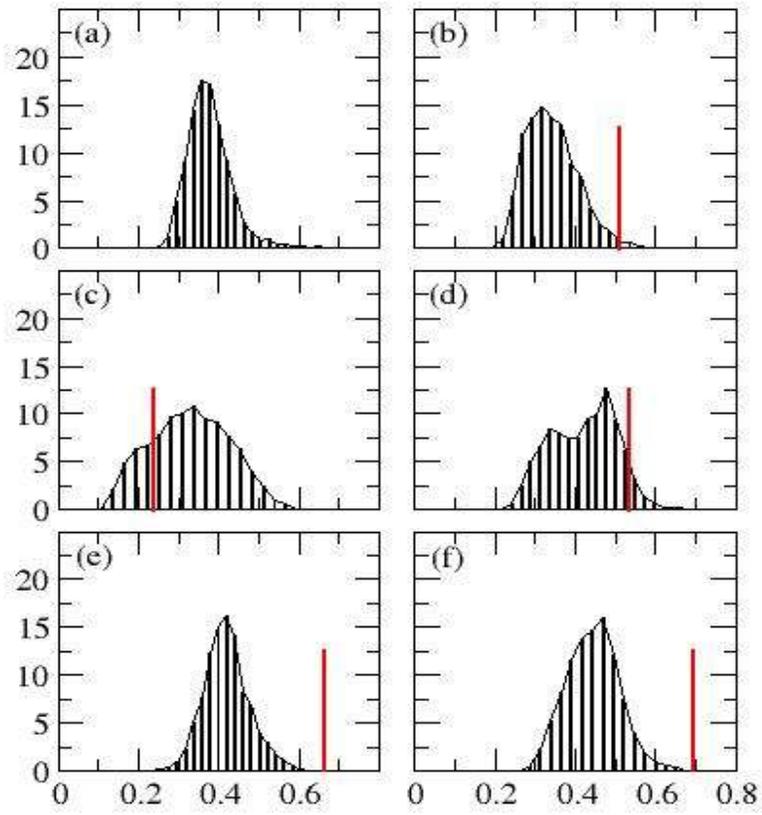

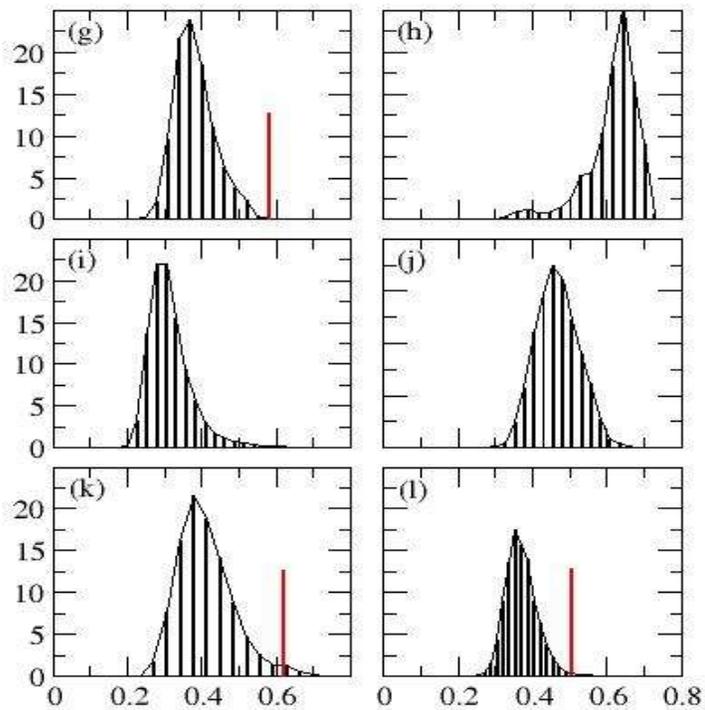

**Figure 2. GC distributions in insect genomes.** Distribution of GC levels of 5,000 random 1Kbase segments from each of the following genomes are shown: (a) *Bombyx mori,* (b) *Tribolium castaneum,* (c) *Apis mellifera,* (d) *Nasonia vitripennis,* (e) *Drosophila melanogaster,* (f) *Anopheles gambiae,* (g) *Aedes aegypti,* (h) *Culex pipiens,* (i) *Acyrthosiphon pisum,* (j) *Ixodes scapularis,* (k) *Homo sapiens* and (l) *S. purpuratus.* The calculations were done on the entire genomes except for *Drosophila* (chr2L) and human (chr14). Red vertical bars show median locations of third codons of protein-coding genes.

(a)

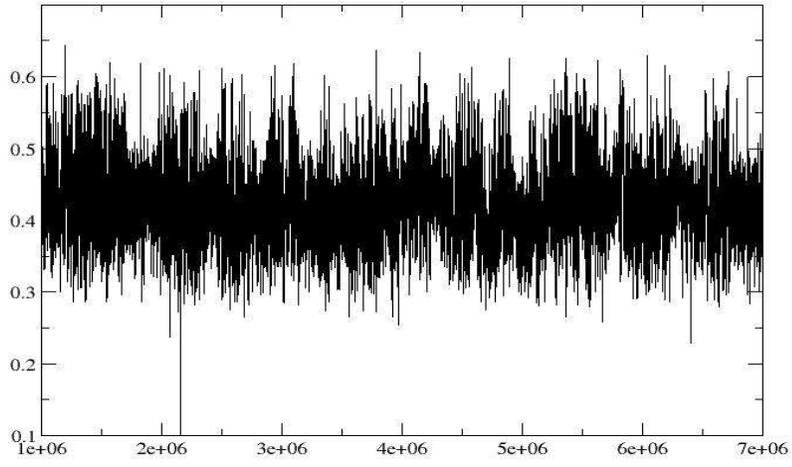

(b)

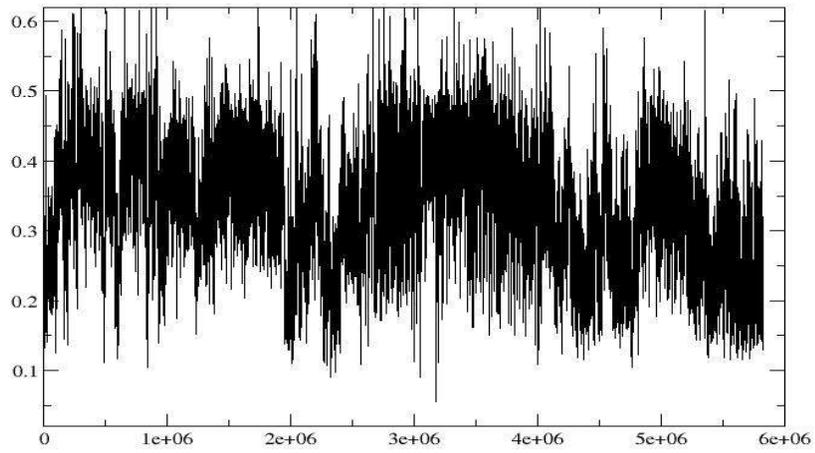

**Figure 3. Variations in bee and *Drosophila* genomes.** GC levels of overlapping windows of 2Kbase sizes, shifted by 1Kbase, were shown for the following genomes: (a) *Drosophila melanogaster*, (b) *Apis mellifera*. If more than 50% of the window was unsequenced, it was not included in the calculation. The images show that the overall variation is larger in *Apis* than *Drosophila*, supporting data in Table 1.

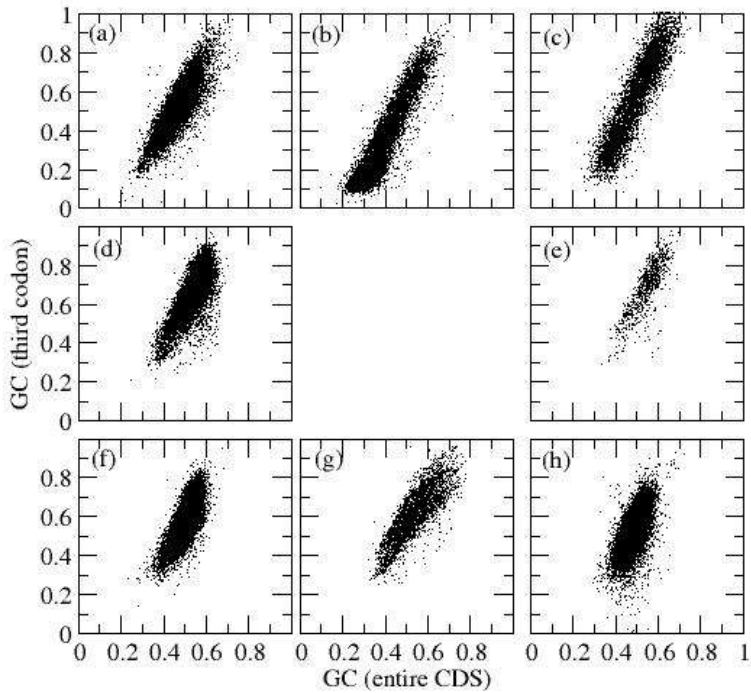

**Figure 4. GC-distributions of third codons.** GC levels of third codons and the entire coding sequences are compared for the following genomes: (a) *Tribolium castaneum,* (b) *Apis mellifera,* (c) *Nasonia vitripennis,* (d) *Drosophila melanogaster,* (e) *Anopheles gambiae,* (f) *Aedes aegypti,* (g) *Homo sapiens* and (h) *S. purpuratus.* They show similar shapes for all insect genomes. The slope in sea urchin is genome is slightly higher.

(a)

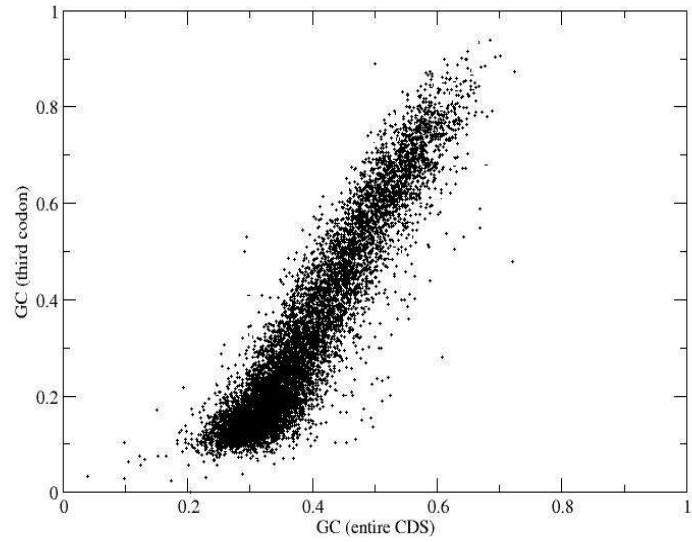

(b)

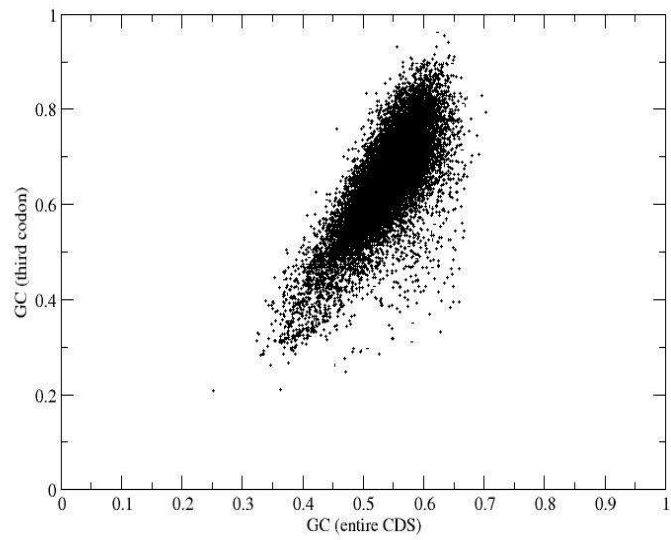

**Figure 5. GC distributions of third codons.** Enlarged versions of images in Fig. 4 for (a) honeybee, (b) *Drosophila* are presented. Most honeybee genes are located in AT-rich

regions of the genome, and therefore the dots for honeybee are present near the bottom left. The distribution is opposite in *Drosophila*, and the most dots are present near top right.